\documentclass[apl,amsmath,amssymb,
reprint 
]{revtex4-1}

\usepackage{graphicx}
\usepackage{dcolumn}
\usepackage{bm}
\usepackage[mathlines]{lineno}

\usepackage[utf8]{inputenc}
\usepackage[T1]{fontenc}
\usepackage{mathptmx}

\begin{document}

\title[Calibration of Cantilevers of Arbitrary Shape]{Determination of Calibration Parameters of Cantilevers of Arbitrary Shape \\by Finite Elements Analysis }

\author{Jorge Rodriguez-Ramos}
\email{jorge.r.ramos@outlook.com}

\author{Felix Rico}
 
\affiliation{ 
Aix-Marseille University, INSERM, CNRS, LAI, \\13009 Marseille, France
}%

\begin{abstract}
The use of atomic force microscopy on nanomechanical measurements requires accurate calibration of the cantilever's spring constant ($k_c$) and the optical lever sensitivity ($OLS$). The thermal method, based on the cantilever's thermal fluctuations in fluid, allows estimating $k_c$ in a fast, non-invasive mode. However, differences in the cantilever geometry  
and mounting angle require the knowledge of three correction factors to get a good estimation of $k_c$: the contribution of the oscillation mode to the total amplitude, the shape difference between the free and the end-loaded configurations, and the tilt of the cantilever respect to the measured surface. While the correction factors for traditional rectangular and V-shaped cantilevers geometries have been reported, they must be determined for cantilevers with non-traditional geometries and large tips. Here, we develop a method based on finite element analysis to estimate the correction factors of cantilevers with arbitrary geometry and tip dimensions. The method relies on the numerical computation of the effective cantilever mass. The use of the correction factor for rectangular geometries on our model cantilever (PFQNM-LC) will lead to values underestimated by 16\%.
In contrast, experiments using pre-calibrated cantilevers revealed a maximum uncertainty below 5\% in the estimation of the $OLS$, verifying our approach.

\end{abstract}

\maketitle

\section{Introduction}

Atomic force microscopy (AFM) has evolved and diversified since its invention in 1986 \cite{Binnig1986}. A mainstream application of AFM is devoted to force spectroscopy measurements to probe the mechanics of materials, including biological systems, such as protein unfolding, receptor-ligand interactions, and the mechanical properties of cells \cite{radmacher1992,moy1994,Radmacher1996,rief1997,Lekka2012,Ramos2014,Hughes2016,Ott2017,Valotteau2019}. 

In a typical AFM setup, a laser beam reflects in the cantilever's back to monitor the deflection from the change in the position of the reflected light on a segmented photodiode. To obtain accurate force measurements, it is crucial to know the conversion factor to transform the electrical signal read in the photodiode (in volts) into the actual displacement (typically nanometers). The more straightforward method to obtain the conversion factor is to deflect the cantilever against a hard surface by obtaining force-distance (FD) curves. Knowing the scanner movement in the vertical direction, the slope of the voltage change of the photodiode provides the conversion factor or optical lever sensitivity ($OLS$), and its inverse ($invOLS= \Delta z / \Delta V$) \cite{Cook2006}.  The $invOLS$ value allows estimating the spring constant by the thermal method from the fundamental mode of oscillation of the cantilever \cite{Stark2001a,Cook2006,Ohler2007,Butt2005,Garcia2010a,Sumbul2020} 

\begin{equation}
    \label{eq:kc}
    k_c = \frac{\beta}{\chi^2}\frac{k_B T}{invOLS^2 \langle V^2 \rangle}
\end{equation}

where $k_B$ is the Boltzmann constant, $T$ is the absolute temperature, $\langle V^2 \rangle$ is the mean-square deflection in volts due to thermal fluctuations of the fundamental mode. The $\beta$ factor corrects the difference between the spring constant of the cantilever (or static spring constant) from that of the fundamental mode $k_1$ (or $k$ dynamic). The factor $\chi$ corrects the difference in the measured deflection of the end-loaded cantilever, respect to the freely oscillating cantilever.  Finally, the cantilever's mean square displacement is  $\langle z_c^2 \rangle =\chi^2 invOLS^2 \langle V^2 \rangle$. Since pushing the cantilever tip against a hard surface is not always possible and may damage the tip, the calibration of both, the spring constant and the $invOLS$ based on the thermal method is becoming popular in biological AFM applications \cite{Butt1995,Cook2006,Higgins2006,Schillers2017,Sumbul2020}.

\begin{figure}
    \centering
    \includegraphics[width=0.45\textwidth]{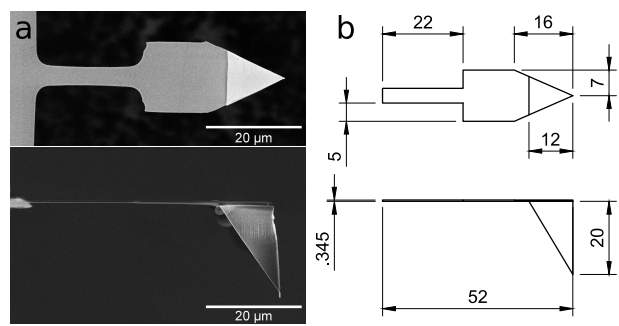}
    \caption{PFQNM cantilever. (a) Scanning electron micrographs of the bottom and lateral views. (b) Relevant dimensions of the simplified geometry used in the simulations by FEA.  }
    \label{fig:fig1}
\end{figure}

The analytical values of $\chi$ and $\beta$, for rectangular cantilevers with a tip of despicable mass and a laser spot infinitely small located at the free end are vastly known  
\cite{Stark2001a,Butt2005,Cook2006,Ohler2007,Ohler2007a,Garcia2010a,Sumbul2020}. However, cantilever geometries are moving towards more irregular shapes. For example, PFQNM-LC-A-CAL cantilevers  (PFQNM hereafter, Bruker)  feature paddle-like geometry to reduce viscous damping  \cite{edwards2017} and have a very large tip, compared to the cantilever size (Fig. \ref{fig:fig1}). The correction factors of PFQNM cantilevers should differ from those of the rectangular beam. On the other hand, manufacturers provide the spring constant's precalibrated values, allowing calibration of the $invOLS$ using either FD curves on a hard substrate or thermal analysis. PFQNM cantilevers feature a large pyramidal tip of $\sim$20 $\mu$m  height with a protruding rounded cylinder of $\sim$70 nm radius, being resistant, in principle, to FC-based calibration. Nonetheless, for cantilevers with tip functionalization or with sharp tips (e.g., PEAKFORCE-HIRS-F family, Bruker), samples placed on a soft surface \cite{solon2007}, or samples covering all the sample surface (e.g., tissue \cite{Lekka2012b,Plodinec2012}, confluent cells or extracellular matrix \cite{Goetz2011}), may not allow FC-based calibration. Thus, it is more convenient to use the thermal method (Eq. \ref{eq:kc}) to calibrate the $invOLS$. This requires accurate knowledge of the correction factors $\chi$  and  $\beta$ for the type of cantilever used. However, analytical expressions for $\beta$ and $\chi$ are only available for rectangular geometries.

There is an important correction to include in Eq.~\ref{eq:kc}. In most AFM experimental setups, the cantilever is mounted at an angle $\theta$ = 10 to 12 degrees with respect to the horizontal plane. It's been noticed that the effective spring constant  of the tilted cantilever ($k_\theta$) is different from $k_c$ (non-tilted). The effective stiffness of a cantilever with a despicable tip mass, loaded at the end will increase by $1/\cos^2 \theta$ \cite{Hutter2005,Ohler2007, Ohler2007a}. When the tip size is large (e.g., PFQNM or colloidal probes), additional corrections are needed \cite{Edwards2008,Chighizola2021}. Although there are analytical expressions for its calculation on rectangular cantilevers, this correction is not well defined for cantilevers and tips of irregular geometry.

\begin{figure*}
    \centering
    \includegraphics[width = .85\textwidth]{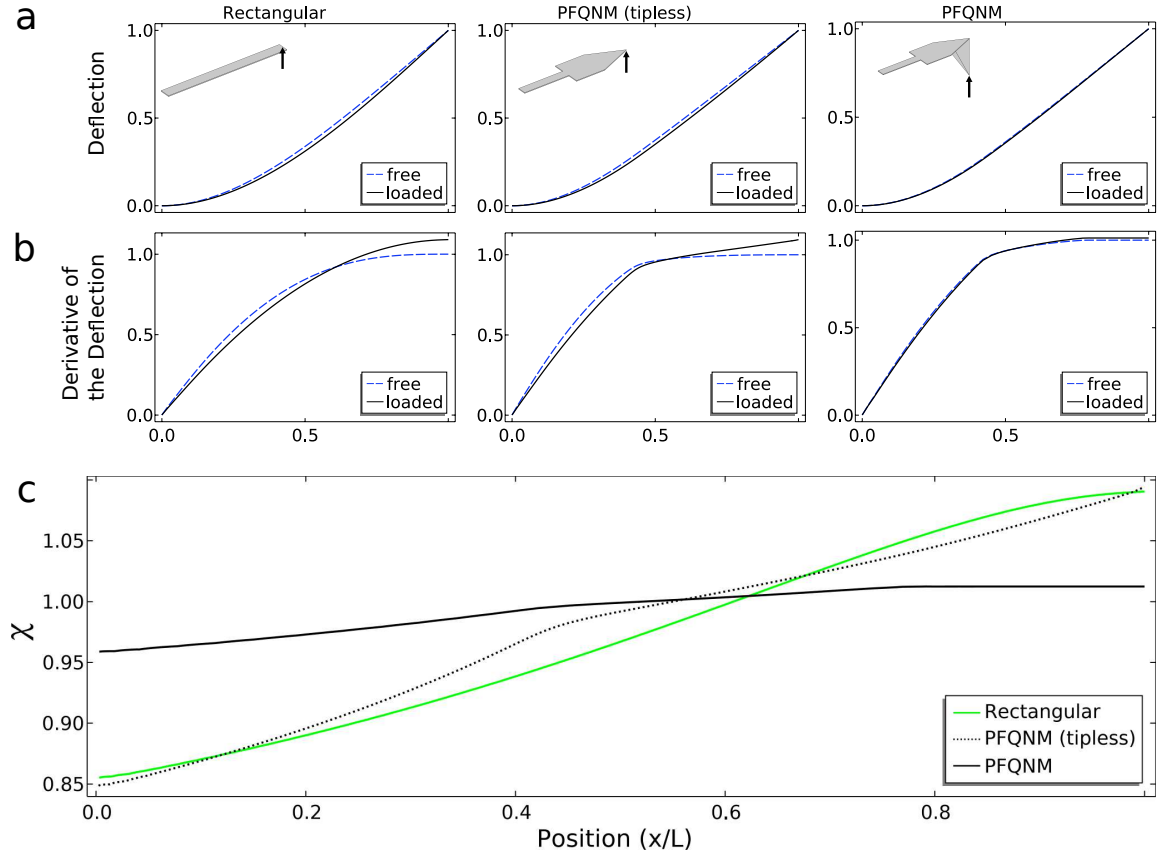}
    \caption{Determination of the correction factor $\chi$ in COMSOL. (a) Deflection of the free and the end-loaded cantilevers, obtained from the eigenmode and stationary studies, respectively. The inset represents the geometry of the cantilever, fixed at $x$ = 0. The arrow represents the point where the load force is applied (stationary study). (b) Derivative of the deflection. (c) Correction factor $\chi$ (Eq. \ref{eq:chi}). See $\chi$ along the cantilever axis for additional geometries in supplementary Fig. S2.}
    \label{fig:fig2}
\end{figure*}

Finite element analysis (FEA) is a common alternative to derive the cantilever's mechanical properties when analytical solutions do not exist. For example, Stark et al. \cite{Stark2001a} used FEA to determine the $\chi$  and  $\beta$ factors from V-shaped cantilevers. While it is relatively simple to extract $\chi$ using FEA, to our knowledge, there is no clear method to determine $\beta$. Here, we implement a method based on FEA to determine the effective mass to calculate $\beta$  for cantilevers of arbitrary shape. Our approach includes the determination of the tilt correction factor and the adjustment of the  manufacturer's pre-calibrated spring constant, to adapt it to our experimental conditions. We apply the method to cantilevers with different geometries and validate it experimentally.

\section{Finite Element Analysis}

We used COMSOL Multiphysics version 5.5  (COMSOL hereafter) to perform the FEA modeling \cite{comsol}. The analysis was applied to a classical cantilever with a homogeneous rectangular section, the simplified PFQNM showed in Fig. \ref{fig:fig1}b and a version of PFQNM without the tip (rectangular, PFQNM and tipless PFQNM cantilevers hereafter). The rectangular cantilever had the dimensions of the PFQNM cantilever provided by the manufacturer in the cantilever’s box: length of 54 $\mu$m, width of 4.5 $\mu$m and thickness 0.345 $\mu$m. The actual dimensions of the PFQNM levers were extracted from scanning electron microscopy micrographs (Teneo VS, FEI) and are shown in Fig. \textbf{1}. For simplicity, the modelled material was silicon with Young’s modulus $E$ = 170 MPa and density $\rho$ = 2329 kg/m$^3$. See supplementary Fig. S1 for geometrical details of the other simulated geometries. To determine both $\beta$ and $\chi$ factors, we run two FEA studies: a static simulation in which a vertical load was applied at tip position (insets, Fig. \ref{fig:fig2}a) to determine the deflection of the end-loaded cantilever, and an eigenfrequency simulation to determine resonance frequency and the modal shape of the free cantilever in vacuum. Here, we will refer to the results of the static and eigenfrequency studies as ‘loaded’ and ‘free’ modelling solutions.

\subsection{Determination of the correction factors $\beta$ and $\chi$}

The factor $\beta$ is defined as

\begin{equation}
    \label{eq:beta}
    \beta = \frac{k_c}{k_1}
\end{equation}

For a rectangular cantilever, $\beta$= 0.971 as determined analytically and from FEA \cite{Cook2006,Garcia2010a,Ohler2007,Butt2005}. In the static study, We obtain $k_c$ by Hooke’s law 

\begin{equation}
    \label{eq:hooke}
    k_c = \frac{F_z}{z_{l}}
\end{equation}

\noindent where $F_z$ is the vertical load applied and $z_{l}$ is the vertical deflection at the point of force application.

In the eigenfrequency study, the fundamental mode of oscillation has a resonance frequency $f_1$ that is related to the dynamic spring constant $k_1$ by

\begin{equation}
    \label{eq:k1}
    k_1 = m_{e}(2 \pi f_1)^2
\end{equation}

\noindent where $m_{e}$ is the effective mass of the cantilever at the resonance frequency, which is 0.25$m_c$ for the rectangular cantilever \cite{Garcia2010a,Hauer2013}, but may differ for a cantilever with irregular geometry and large tip dimensions. We used the method proposed by Hauer et al. \cite{Hauer2013} to define the effective mass integral 

\begin{equation}
    \label{eq:meff}
    m_{e}(x_{l}) = \frac{1}{|r_1(x_{l})|^2}\int_V dV \rho(x) |r_1(x)|^2
\end{equation}

\noindent where $\rho$ is the density and $r_1$ is the first mode shape solution along the longitudinal axis ($x$); $x_l$ represents the position where the load $F_z$ is applied (i.e., at the tip position). This approach was validated by the calibration values obtained for V-shaped cantilevers (Table \ref{tab:sim}), where the force is not applied at the free end. Indeed, the work of Hauer \cite{Hauer2013} suggests that the effective mass is a function of the position at which we measure the device; in our case, at the point where the load is applied. Importantly, Eq.~\ref{eq:meff} applies to geometries of arbitrary shape and we solved it numerically for the different cantilever geometries using COMSOL. As expected, we obtained the same analytical value of $m_e$ of 0.25 of the cantilever mass ($m_c$) for the rectangular cantilever.

\begin{figure*}
    \centering
    \includegraphics[width=.85\textwidth]{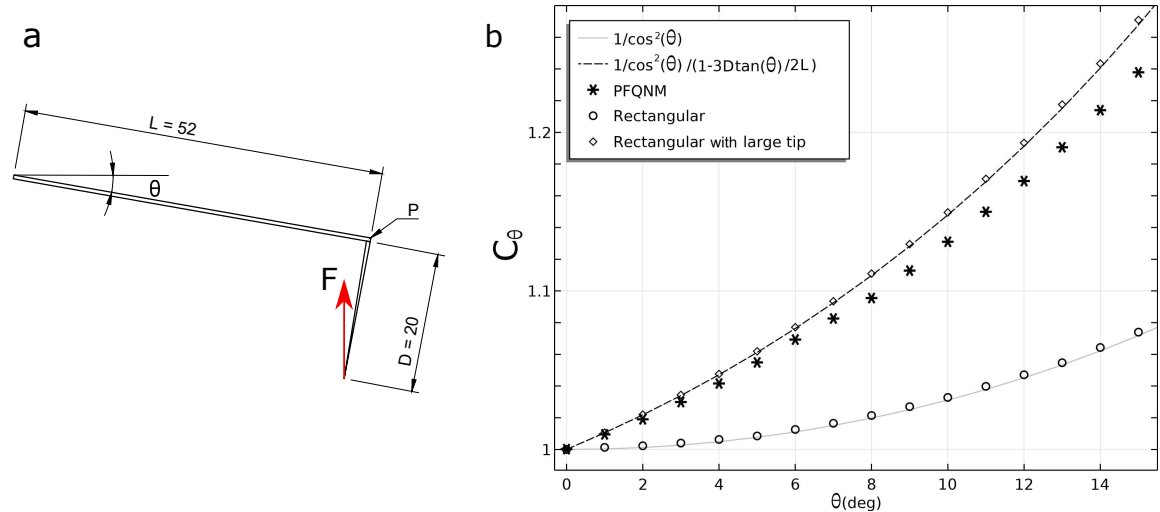}
    \caption{Cantilever tilt correction factor $C_\theta$ (Eq.~\ref{eq:kc_ratio}). (a) Schematic of a  rectangular cantilever  with a long sharp tip, with the same tip height and cantilever length as the PFQNM modelled in this work (Fig.~\ref{fig:fig1}a). The force $F$ is applied at the very end. The spring constant is determined at point P, located within the cantilever body. (b) FEA calculated values for the classical rectangular cantilever (tip-less), the rectangular cantilever with idealised long sharp tip (a) and the PFQNM cantilever. The gray solid line represents the theoretical value for a rectangular cantilever $1/\cos^2 \theta$, while the black dashed line is the theoretical value for a cantilever with a sharp large tip (Eq.~\ref{Eq:edwards}).}  
    \label{fig:tilt}
\end{figure*}

To obtain $\chi$, we determined the derivative the deflection along the cantilever longitudinal axis of the loaded and free solutions and computed their quotient

\begin{equation}
    \label{eq:chi}
    \chi= \frac{invOLS_{free}}{invOLS}=\frac{\frac{d}{dx} (z_{loaded})}{\frac{d}{dx} (z_{free})}
\end{equation}

\noindent where $z_{loaded}$ and $z_{free}$ are the vertical deflection of the loaded and free solutions, respectively. For a rectangular cantilever, we obtained $\chi$ = 1.09, as reported before, using analytical and FEA approaches \cite{Cook2006,Garcia2010a,Ohler2007,Proksch2004} (Table \ref{tab:sim}).

\subsubsection{PFQNM cantilevers}

The values of $m_e$, $\beta$ and $\chi$ obtained for different cantilever geometries are shown in Table \ref{tab:sim}. We also report, as reference, the values for geometries  reported in the literature that confirm our approach (such as V-shaped MLCT-D, Bruker and arrow-shaped AC160TS, Olympus)\cite{Stark2001a,Sader2014b}. As shown in Fig. \ref{fig:fig2}a, b, for the rectangular and the tipless PFQNM cantilevers, the deflection of the freely oscillating cantilever is different from the deflection of the end-loaded; the same applies for the deflection's derivative. However, this difference is less pronounced in the case of the PFQNM (including the tip). This suggests that the mass of the PFQNM cantilever tip (which accounts for approximately 60 to 75\%  of the cantilever’s total mass, see supplementary Table S1) has a strong influence in its mechanical behaviour. The consequence is a factor $\chi$ much closer to 1 for the PFQNM cantilever with respect to the other two cantilever types (Fig. \ref{fig:fig2}c and Table \ref{tab:sim}). In addition, the $\chi$  factor remains approximately the same for the last 20\% of its length towards the free end. This implies that force measurements will be less affected by little changes in the position of the laser spot.

\begin{table}[ht]
    \caption{Estimated parameters of selected cantilevers.}
    \label{tab:sim}
    \begin{ruledtabular}
    \begin{tabular}{lcccc}
        Cantilever     & $m_e/m_c$&  $\beta$   & $\chi$\footnote{$\chi$ value at the end of the cantilever.}  
        & $\beta/\chi^2$ \\
        \hline
        Rectangular, analytical \footnote{See references \cite{Cook2006,Garcia2010a,Ohler2007,Butt2005}.}
        & 0.250     & 0.971     & 1.090	    & 0.8175 \\
        Rectangular      	    & 0.249	    & 0.971	    & 1.090	    & 0.8173 \\
        \\
        PFQNM (tipless)         & 0.264	    & 0.966	    & 1.094	    & 0.8068 \\
        PFQNM \footnote{See supplementary Table S1 for simulations considering a the reflective gold coating on the cantilever's back.}                  
        & 0.631	    & 0.998	    & 1.012	    & 0.9737 \\
        \\
        MLCT-E like, Ref. \cite{Stark2001a}
                                & --        &0.963	    & 1.125	    & 0.7608\\
        MLCT-E, V-shaped        & 0.232	    & 0.956	    & 1.125	    & 0.7554\\
        MLCT-D, V-shaped        & 0.227	    & 0.959	    & 1.116	    & 0.7690 \\
        \\
        AC160TS, Ref. \cite{Sader2014b} 	
                                &-- 	    &0.908	    &1.254	    &0.5776\\
        AC160TS (tipless)	    & 0.151	    & 0.904	    & 1.271	    & 0.5600\\
        AC160TS	                & 0.156	    & 0.915	    & 1.217	    & 0.6177\\
    \end{tabular}
    \end{ruledtabular}

\end{table}

Despite obvious geometrical differences, the values of $\chi$ and $\beta$ for the tipless PFQNM cantilever are within $\sim$0.5\% from the rectangular ones (Table \ref{tab:sim}). However, the $\chi$  and $\beta$ values for the PFQNM cantilever, including its large tip, differ importantly from the rectangular ones; the PFQNM values are  7.7\% lower ($\chi$) and 2.8\% higher ($\beta$). Thus, using the rectangular correction factors to calibrate the $invOLS$ from Eq. \ref{eq:k1} on a PFQNM cantilever will lead to a non-negligible $\sim$16\% error in the estimation of the $invOLS$ and, subsequently, in the determination of the measured forces.

\subsubsection{V-shaped cantilevers}

Regarding V-shaped cantilevers, the correction factors $\chi$ and $\beta$ were estimated by Stark et al. \cite{Stark2001a} for, at the time, the Thermomicroscopes type E cantilever, with similar dimensions to the MLCT-E from Bruker (Table \ref{tab:sim}). Our simulated values are very close to those reported before and are similar (within $\sim$1\%) between MLCT-D and E. However, it is important to note that the calibration parameters will depend on the position of the tip. Even for MLCT cantilevers within the same chip, where the distance to the tip-end is the same (in our case, 7 $\mu$m), the relative position of the tip will be different for different cantilevers dimensions (A to F). 

\subsubsection{AC160TS cantilevers}
AC160TS cantilevers, which have been simulated by FEA before \cite{Sader2012,Sader2014}, constitute another interesting example to assess the importance of the tip mass. Our simulations of an AC160TS cantilever without the tip, are close to the values reported in the literature (Table \ref{tab:sim}). However, when we include the tip on the simulation, the correction factor $\chi/\beta^2$ increases by 10\%. Overall, the tip of the AC160 is almost as large as that of the PFQNM (see supplementary information), but its influence is smaller because its mass represent only 1\% of the total mass.

It is important to note that the $\chi$ values shown in Table \ref{tab:sim} correspond to the end of the cantilever. However, it is common to place the laser spot before the end.  This translates into a $\chi$ smaller than $\chi$ at the end \cite{Proksch2004,Sader2014}. We show the $\chi$ values along the $x$ axis of the cantilevers in Fig. S2.

\subsection{Cantilever tilt}

In   general, the relationship between the spring constant of the non-tilted cantilever $k_c$, and the effective spring constant $k_\theta$ of the same cantilever, mounted with an angle $\theta$ in the AFM system is

\begin{equation}
    \label{eq:kc_ratio}
    k_\theta = C_\theta k_c
\end{equation}

\noindent where the factor $C_\theta$ will depend on the tilt angle and the tip's geometry and position \cite{Edwards2008}.

By substituting Eq.~\ref{eq:kc_ratio} in Eq. \ref{eq:kc} we get an equation for the tilted cantilever

\begin{equation}
    \label{eq:kc*}
    k_\theta = \frac{\beta}{\chi^2}\frac{C_\theta}{invOLS^{2}} \frac{k_B T}{ \langle V^2 \rangle}
    =\beta C_\theta \frac{k_B T}{\langle z_c^2 \rangle}
\end{equation}

If we define the effective $invOLS$ of the tilted cantilever as 

\begin{equation}
\label{eq:invOLS*}
invOLS_\theta=\frac{invOLS}{\sqrt{C_\theta}}
\end{equation}

\noindent we can rewrite Eq.~\ref{eq:kc*} to estimate the effective spring constant when using the $invOLS_\theta$ from static force curves, without need to correct the tilting or the cantilever geometry

\begin{equation}
    \label{eq:kc*_invOLS*}
    k_\theta = \frac{\beta}{\chi^2}\frac{k_B T}{invOLS_\theta^{2} \langle V^2 \rangle}
\end{equation}

\subsubsection{Determination of the tilt correction factor $C_\theta$ by FEA}

We need to implement Eq.~\ref{eq:kc_ratio} to determine the correction factor $C_\theta$  by FEA. Figure~\ref{fig:tilt}a shows a rectangular cantilever with a large tip. The lever is tilted an angle $\theta$ respect to the horizontal. The force is applied at the very end of the cantilever. The spring constant is determined by the Hooke's law (Eq.~\ref{eq:hooke}) in the point P (perpendicular to the point of application of the force in the reference system of the cantilever). The theoretical $C_\theta$ value for the  cantilever-tip system depicted in Fig.~\ref{fig:tilt}a  \cite{Hutter2005,Edwards2008} is

\begin{equation}
    \label{Eq:edwards}
    C_\theta = \left[ \cos^2\theta \left( 1-\frac{3D}{2L}\tan \theta \right) \right]^{-1}
\end{equation}

\noindent where $D$ is the tip height and $L$ the cantilever's length.

Figure~\ref{fig:tilt}b shows the simulated values of $C_\theta$ for three geometries. There is a very good agreement between the simulated  and the theoretical value for the bare rectangular cantilever ($D=0$). The same occurs for the rectangular cantilever with the long tip and the theoretical value from Eq~\ref{Eq:edwards}. However, $C_\theta$ for the PFQNM cantilever deviates from Eq~\ref{Eq:edwards}, even though the ratio $D/L$ is the same as for the idealized cantilever with long tip in Fig~\ref{fig:tilt}a. Possible reasons for this difference are the paddle geometry of the cantilever or the larger zone of interaction between cantilever and the base of the pyramidal tip.

\subsubsection{Practical implementation using manufacturer pre-calibrated cantilevers} 

The spring constant of PFQNM probes is pre-calibrated by the manufacturer. The mean square displacement $\langle z_c^2 \rangle$ of each cantilever is determined by using a laser Doppler vibrometer (LDV) as described by Ohler \cite{Ohler2007}. For simplicity, the manufacturers do not consider the tip height in the calibration process (i.e., $D$ = 0 in Eq.~\ref{Eq:edwards}). In that case, the equation used to determine the pre-calibrated spring constant $k_\mathrm{cal}$ is

\begin{equation}
    \label{eq:kcal}
    k_\mathrm{cal}=
    \frac{\beta} {\cos^2 \theta}\frac{k_B T}{ \langle z_c^2 \rangle} = 1.0149 \frac{k_B T}{ \langle z_c^2 \rangle}
\end{equation}

\noindent where $\beta=0.971$ (rectangular cantilever) and  $\theta = 12^{\circ}$, the most common tilt angle in  Bruker AFM systems.

Since the mean square displacement of the cantilever $\langle z_c^2 \rangle$ should be the same regardless the measuring technique, we can combine Eq.~\ref{eq:kc*} and Eq.~\ref{eq:kcal} to obtain a `corrected' value of the pre-calibrated spring constant for our AFM system

\begin{equation}
    \label{eq:kcorr}
    k_\mathrm{corr}=\frac{\beta C_\theta}{1.0149} k_\mathrm{cal}
\end{equation}

Note that for a rectangular cantilever with tip of despicable mass $k_\mathrm{corr} = k_\mathrm{cal}$. In our experimental conditions, we used  PFQNM cantilevers $\beta=0.998$ (Table~\ref{tab:sim})  mounted in a Nanowizard 4 AFM system with  $\theta = 10^\circ$, then $C_\theta=1.131$ (Fig.~\ref{fig:tilt}b), and  $k_\mathrm{corr} = 1.1122 k_\mathrm{cal}$ (i.e., the effective stiffness of PFQNM cantilevers is expected to be $\sim$11\% higher than the calibrated value provided by the manufacturer). 

The effective $invOLS$ based on the pre-calibrated spring constant provided by the manufacturer will be
\begin{equation}
    \label{eq:invOLS*_kcal}
    invOLS_\theta = \sqrt{ \frac{1.0149}{ \chi^2 C_\theta}
    \frac{k_B T}{ k_\mathrm{cal} \langle V^2 \rangle}}
\end{equation}

\subsubsection{Practical implementation using cantilevers calibrated by Sader's method}

The Sader method is widely used to determine the spring constant of a large number of commercial cantilevers \cite{sader2016,Higgins2006,Sader2012}. In this case, the effective $invOLS$ would be determined by

\begin{equation}
    \label{eq:invOLS*_sader}
    invOLS_\theta = \sqrt{ \frac{\beta}{ \chi^2}\frac{k_B T}{C_\theta k_\mathrm{Sader} \langle V^2 \rangle}}
\end{equation}

\begin{figure}
    \centering
    \includegraphics[width=.45\textwidth]{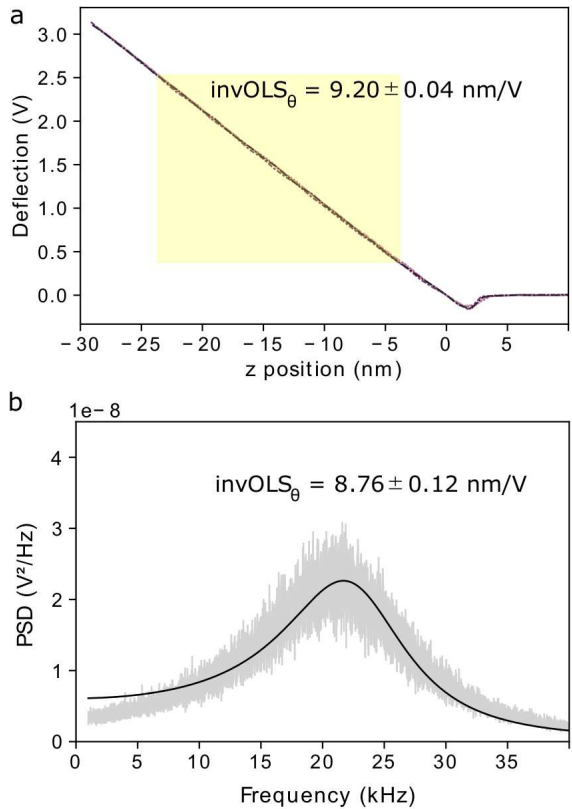}
    \caption{Experimental determination of the effective $invOLS$ of cantilever No. 1 in Table \ref{tab:exp}. (a) Overlap of five force curves on a rigid surface, in liquid media. The $invOLS_\theta$ is the inverse of the slope of the linear region of the force curves. The fitting was performed in the center of the zone of interest, between 0.5 and 2.5 V (shaded region). (b) $invOLS_\theta$ estimation from thermal spectra in liquid (Eq. \ref{eq:invOLS*_kcal}),  ($n$ = 5).}
    \label{fig:fig3}
\end{figure}

\begin{table*}[!t]
    \caption{Experimental determination of the effective $invOLS$ and effective spring constant of PFQNM cantilevers.\footnote{Values reported as mean $\pm$ standard deviation.}}
    \label{tab:exp}
    \begin{ruledtabular}
    \centering
    \begin{tabular}{cccrc}
    Lever & \multicolumn{2}{c}{$invOLS_\theta$ (nm/V)} &
    \multicolumn{2}{c}{ $k_\theta$ (mN/m)} \\
    No.               
    & static \footnote{Based on force curves in rigid surface (Fig. \ref{fig:fig3}a). }
    & thermal (Rel. error) \footnote{Based on thermal noise spectra (Fig. \ref{fig:fig3}b, Eq.~\ref{eq:invOLS*_kcal}).}     
        
    &  corrected \footnote{Corrected value of the manufacturer's spring constant (Eq.~\ref{eq:kcorr}).}    
    & estimated (Rel. error) \footnote{Equation~\ref{eq:kc*_invOLS*}.} \\
    
    \hline
    1	& 9.20 ± 0.04	& 8.76 ± 0.12	(-4.8\%)	    & 129	    & 117 ± 3	    (-9.3\%) \\
    2	& 9.12 ± 0.31	& 8.96 ± 0.06	(-1.7\%)	    & 107	    & 103 ± 1	    (-3.7\%) \\
    3	& 9.12 ± 0.20	& 8.68 ± 0.03	(-4.9\%)	    & 103	    & 94 ± 1	    (-8.7\%) \\

    \end{tabular}
    \end{ruledtabular}

\end{table*}

\section{Experimental validation}

To further verify our FEA approach, we experimentally determined the $invOLS$ using both thermal and FC-based methods on three PFQNM cantilevers with pre-calibrated spring constant (Bruker). Experiments were performed in a commercial AFM system (JPK Nanowizard 4, Bruker). For each cantilever, three thermal spectra were recorded in liquid media (10 mM Tris, 150 mM KCl, pH 7.4, Merck), keeping the cantilevers more than 500 $\mu$m away from a freshly cleaved mica surface. Then, five force curves were acquired on the mica surface to a force setpoint of 3 V, which guaranteed to have a sufficiently wide linear region to determine the $invOLS$ around the center of the photodiode Fig. \ref{fig:fig3}a. Finally, three new spectra were recorded far from the surface. The thermal spectra were fitted with the damped simple harmonic oscillator (SHO) model 

\begin{equation}
    \label{eq:sho}
S = A_w^2 + \frac{A^2f_1^4}{Q^2}\left[ (f^2 - f_1^2)^2 + \frac{f^2 f_1^2}{Q^2} \right]^{-1}
\end{equation}

\noindent where $A_w$ is the background noise, $A$ is the amplitude at the resonance frequency ($f_1$), and $Q$ is the quality factor (Fig. \ref{fig:fig3}b). Then, we calculated the mean-squared deflection in volts as \cite{Sumbul2020} 

\begin{equation}
    \label{eq:d2}
    \langle V^2\rangle = \frac{\pi A^2 f_1}{2Q}
\end{equation}

The $invOLS_\theta$ was recovered by substituting Eq.~\ref{eq:d2} into Eq.~\ref{eq:invOLS*_kcal}, using the pre-calibrated spring constant. The results are summarized in Table \ref{tab:exp}. Even if the five force curves were performed micrometres away from each other, they were almost indistinguishable when aligned to zero force (Fig.~\ref{fig:fig3}a), resulting in $\sim2\%$ average uncertainty. Overlapping was also observed in the thermal spectra (Fig. \ref{fig:fig3}b), even if those were recorded before and after the force curves acquisition, resulting in even lower uncertainty $<1\%$. 

For the three cantilevers measured, we found a good agreement between the $invOLS_\theta$ derived from the force curves and from the thermal spectra. Note that, since we have the calibrated value from the manufacturer, the factor $\beta$ does not appear in Eq.~\ref{eq:invOLS*_kcal}. We obtained a relative difference  $\sim 4$\% on average, far below the 16\% uncertainty using the correction $\chi$ and $\beta$ for rectangular cantilevers. We obtained SEM images of all three PFQNM cantilevers and tips after the AFM experiments to accurately determine their geometry, finding small variability between cantilevers and tips (Fig. S3).

The comparison of the `corrected' value of the spring constant with the calculated  stiffness  using $invOLS_\theta$ from static force curves (Eq.~\ref{eq:kc*_invOLS*}) leads to a fairly good  agreement (the  average is $\sim7$\%, Table \ref{tab:exp}). As has been shown before, these results suggest that the thermal determination of the $invOLS$ leads to less uncertainty (1/2) compared to the classical approach \cite{Schillers2017}.

If the experiments are performed under optimal conditions on a clean hard surface, and the vertical movement of the piezo-scanner is well calibrated, the actual $invOLS_\theta$ value is the one obtained from slope of the force curves. The difference in values could be due to the influence of the reflective coating on the cantilever mass (see supplementary Table S1), small geometrical differences between cantilevers of different batches,  the calibration of the piezo-scanner or the values reported by the manufacturer; these two last factors were considered to be error-free in the calculations shown in Table~\ref{tab:exp}. 

In summary, we implemented a finite element analysis method to determine the correction factors $\chi$, $\beta$ and $C_\theta$ to calibrate the spring constant and the $invOLS$ by the thermal tune method. Our simulations agree with the values reported in classical geometries, like the rectangular and the V-shaped, which suggests the method is valid for arbitrary shape. As relevant case within non-standard geometries, we focused on PFQNM cantilevers, characterized by a paddle shape with a large tip relative to the cantilever dimensions. We found that a beneficial effect of the massive tip is the little variation in the correction factor $\chi$ near its free end. Thus, laser positioning changes will produce a little variation of $\chi$, resulting in more robust experimental results. In addition to that, our method provides a `correction' to the pre-calibrated value provided by the manufacturer. We found a good agreement between the effective $invOLS$  obtained from force curves and the thermal tune method in AFM experiments with PFQNM cantilevers, confirming the approach's validity.

\section*{Supplementary material}
See supplementary material for additional cantilever geometries,  correction factors $\chi$ for the cantilevers studied and additional simulations of PFQNM cantilevers, considering the reflective gold layer coating.  

\begin{acknowledgments}
We thank Alexander Dulebo (Bruker) for kindly providing technical information and Alessandro Podestà and Matteo Chighizola for insightful discussions. The electron microscopy experiments were carried on the PICsL-FBI core electron microscopy facility (Nicolas Brouilly, IBDM, AMU-Marseille UMR 7288), member of the national infrastructure France-BioImaging supported by the French National Research Agency (ANR-10-INBS-0004). This project has received funding from the European Research Council (ERC, grant agreement No 772257). 
\end{acknowledgments}

\section*{Data availability}
The data that support the findings of this study are available from the corresponding author upon reasonable request.

\bibliography{main}

\end{document}